# On the Current Measurement Practices in Agile Software Development

Taghi Javdani[1], Hazura Zulzalil[1], Abdul Azim Abd Ghani[1], Abu Bakar Md Sultan[1]

[1] Faculty of Computer Science and Information Technology, University Putra Malaysia (UPM)
Serdang, Selangor 43400, Malaysia
tjavdani@yahoo.com, {Hazura,azim,abakar}@fsktm.upm.edu.my

**Abstract**

Agile software development (ASD) methods were introduced as a reaction to traditional software development methods. Principles of these methods are different from traditional methods and so there are some different processes and activities in agile methods comparing to traditional methods. Thus ASD methods require different measurement practices comparing to traditional methods. Agile teams often do their projects in the simplest and most effective way so, measurement practices in agile methods are more important than traditional methods, because lack of appropriate and effective measurement practices, will increase risk of project. The aims of this paper are investigation on current measurement practices in ASD methods, collecting them together in one study and also reviewing agile version of Common Software Measurement International Consortium (COSMIC) publication.

## 1. Introduction

Agile software development (ASD) methods in last decade were introduced as a reaction to traditional software development methods. Emphasize on new values leads these methods to provide different processes for software development. For example handling user requirement change, within all stages of software production, is a principle in ASD [1], thus software requirement management process is completely different from respective process in traditional methods. Since measurement and estimation practices defined based on the processes, it seems that agile measurement practices are different from traditional methods. Also lack of comprehensive documentation in ASD methods, causes some of the famous and popular measurement practices in traditional methods are not usable in ASD methods [2]. Another issue is that, because of emphasize on acceptation of unpredictable change requests in ASD methods, using appropriate and acceptable measurement is a significant necessity in these methods; at least for cost estimation which is the most important factor in point of view of managers. However, there are many standards and well-known measurement practices in traditional methods, there are only a few methods for measuring in ASD methods. Focusing on software work as a value in ASD methods, leads little attention on other activities and practices, but measuring in software development has not only a significant role in creating of value in software development but also, can monitor and reduce risk of the development.

Until late of 2011, there was not an official publication for measurement in ASD methods. At that time, COSMIC introduced first official guideline in ASD methods for software sizing based on function point.[3]

However there are some studies on some specific measurement practices in agile, but there is no study that collect all of them in a paper and explain them in one study, hence we have tried to collect and summarize all agile measurement practices in this paper.

The rest of this paper is organized as follows. Section 2 provides nature of ASD methods, section 3 provides investigation on current agile measurement practices, section 4 explains the new COSMIC guideline and section 5 provides summer and conclusion.

## 2. Agility

### 2.1 Agile history

In 2001, seventeen of the agile advocators and leaders came together and introduced formally ASD principles as "agile manifesto" in software development industry [1]. Although agile concepts and practices were not new at that time [4], it was the first time that a formal definition of agile practices and principles were introduced in software industry. However, fist reactions of proponents of traditional methods were cautious, but when Boehm published his idea on this innovative approach [5], many software practitioners were interested in agile. Currently, many agile methods are popular with some of them being focused on software development, whereas others target on project management. In last decade and during maturing of agility in software industry, many companies and individuals have migrated to these methods [6, 7].

2.2 Agile Principles

Initial ideas of agile are simple and valuable thus no one could reject them or disagree with them. Agile manifesto was based on those simple ideas[1]. reading the manifesto give us these idea clearly:

> "We are uncovering better ways of developing software by doing it and helping others do it. Through this work we have come to value:
> • Individuals and interaction over process and tools,
> • Working software over comprehensive documentation,
> • Customer collaboration over contract negotiation,
> • Responding to change over following a plan.
> That is, while there is a value in the items on the right, we value the items on the left more."

Manifesto and its subsequent notes, explain value principles and ideas in agile, [8-11] in the following items: people oriented, embracing changes, focusing on product, simplicity, self-organized team, fast delivery, iteration, improve quality continually and etc. Based on these ideas different agile methods have been introduced; some of them are development-based and others are project-based.

## 3. Current Agile measurement Practices

ASD methods based on their values and principles prepare less detailed or useful advices about processes. A quick view in planning, sizing and effort estimation as well as project management shows that such activities are really light-weight. In this study we don't focus on any particular agile method and hence our discussion is on common and well-known definition of agile methods.

Most of the measuring practices in traditional software development methods are not usable in agile methods directly, but, tailored methods might be applicable. The current practices of agile measurement could be categorized in the seven classifications, as described in the following subsections.

3.1 Effort estimation, Software size

Using popular effort estimation measurements in software industry needs standard and well-defined user requirement document. In ASD methods not only are there no such detail documents, but also requirements are subject to change [12-16]. User requirements in agile methods are defined as user stories (US) and are collected in backlog. The popular and most common approach for effort estimation in agile methods is subjective estimation [16]. Although this approach is simple and easy to apply, estimates are highly biased [13]. In some agile teams, effort estimation is based on their previous iteration actual effort and hence effort estimation is useful only for remained user stories. In addition application of planning poker is one the most popular practices for many agile teams in planning and predicting effort before starting each iteration [17]. In this method, almost all developers collaborate in estimation, thus, no one estimates for all and also, every one estimates often. In this Practice, each member gives a point to a story and the final point of that story is the mean of its assigned point. Nonetheless, User Story Point (USP) is not objective [18] and cannot define a standard practice for estimation of software size and complexity. More ever there is no evidence that estimation in this way is more adequate than the famous Wideband Delphi [19], but at least it is funny for the teams and motivates them in estimation practice. There are also many reports about using usual software measurements practice which were used in some specific agile methods such as scrum and XP with appropriate customizations [20-22]. Abrahamsson in his latest paper on agile prediction [23], has described a better method for effort estimation in agile. His method relies on predictors extracted from completed user stories and will be used for next stories. He proposed a model for effort estimation to which the effectiveness of the model is different from case to case and is based on quality and style of user stories.

3.2 Velocity (Productivity)

'Velocity' is used in agile communities instead of 'productivity'.[2] Velocity is defined as number of completed user stories in iteration and is often used for estimating of remained time to end of project (However, by this definition, velocity does not mean productivity). This measurement is almost so useful for stable agile teams that are comprised of the same individuals working full time. Since in an iteration all tasks will be done for user stories (designing, Implementing, testing integrating of UI, databases, forms, reports), velocity could be a good measurement for overall productivity of teams, but, a clear threat is using short-term velocity instead of sustainable velocity. Measuring velocity should be done for several successive fix length iterations. This is the only way that velocity measurement could be useful in prediction (see figure 1). Furthermore, any changes to team composition will invalidate the Velocity measurement. Note that, even adding high skilled people to a team can affect on velocity and may reduce it. In fact, velocity relies on team consistency in order to be most valuable, but note that past performance does not guarantee the future results [24]. Velocity is an empirical observation and is not and estimation or target to aim for it. [24].

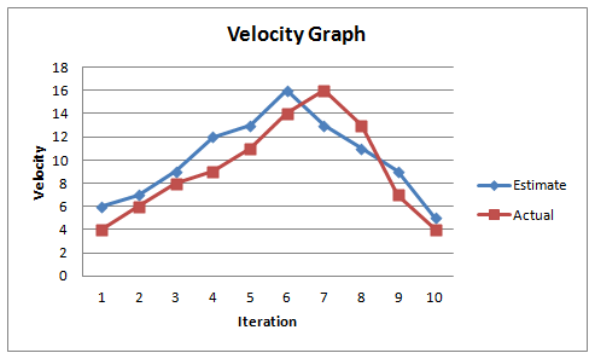

Fig. 1 Velocity graph in agile

Figure 1 shows velocity of an agile team in 10 iterations. Estimated line shows velocity of team based on previous projects and actual line illustrates real velocity of team during current project.

### 3.3 Burndown Chart

Burndown chart is an important measurement tool for planning and monitoring of progress in agile methods based on software working as an agile principle [25]. In most of methods and among many of teams, it is used for representation of amount of remained work [2]. This chart is commonly used in two different types: *iteration burndown* and *release burndown*; for estimating of the remaining work of the tasks to be completed in the iteration and for current release respectively.
A comparison between estimated work (ideal burndown) and remained work together could help teams for decision making in adding or dropping some user stories in case of project is ahead or behind of schedule. Burndown chart is not designed to produce an accurate report on progress of a user story daily. But it is a view of amount of the remained work and if the team based on its velocity is able to achieve its goals in the iteration [2] This is a high level visibility of this tool. Burn up chart is an alternative chart which is used instead of burndown chart. It relies on this concept that an agile project has an unknown size, so, it focuses on incremental progress and work done instead of work remained. Either chart have significant role for controlling of project in overall project progress and not in specific time within the project.

### 3.4 Cumulative Flow

Cumulative flow diagram (CFD) was introduced by Anderson in 2003 as a better replacement for burn up chart [26]. It presents a quantity of work in a given state. Figure 2 shows an example of CFD. In this figure quantity of work in each iteration, is illustrated in one or more predefined states (Design, Development, Test and Deployed)

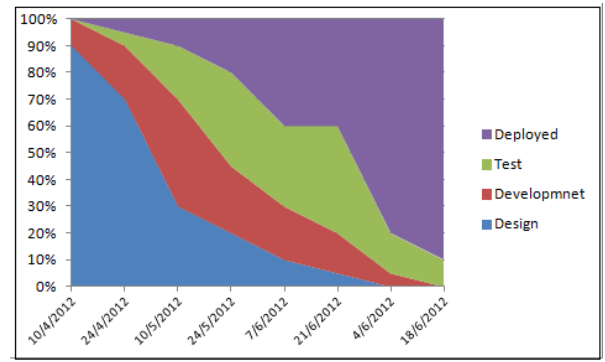

Fig. 2. Cumulative Flow Diagram/Graph

Using CFD is not only useful for progress controlling by depicting work in progress (WIP), but also for increasing throughput and also reducing lead-time for attaining high responding to customer requirements [27]. By this idea, WIP can predict and estimate in advance the delivery date and lead time [28, 29]. Thus, it can be used to correct problems before they become too crucial. Since postponing measuring delivery date, may be causes greater problems in projects, CFD can be useful for solving this issue.

Other benefits of CFD are monitoring bottlenecks in software development work flow [30] and achieve continuous improvement by eliminating bottlenecks [26]. It also shows whether the scope of a project is changing and answers this important question: Is the total size of the backlog (the sum of all of the scope, regardless of status) constant, increasing, or decreasing?

To sum up, the cumulative flow diagram enables teams to measure how efficiently they are delivering valuable, working product to the customer, and indicates where they need to focus their process improvement efforts. Tracking and monitoring of a project with a CFD is a key factor in moving to a implementing a Lean and agile system for software development [26].

3.5 Responding to Change/Re-work

Embracing change is one of the agile principles [1]. This flexibility gives customers opportunity to change their requirement to achieve a better product. Each change could be as a new user story and also as a part of implemented portion. ('Re-work')

Many teams, use re-work as an indicator of ability of team to hand over product quality. From perspective of a metric, re-work can be defined as the man-hours spent fixing flaws and defects. This can be compared to amounts of man-hours spent for developing new tasks. Re-work measurement within each iteration, is helpful to ensure it does not get out of control [31]. One way for demonstration of re-work is using re-work graph, which is also useful for discovering bottlenecks and delays on project. However, many reports claim that cost of changes in agile is too lower compare to other methods [32-34]; but, it is still important for project managers.

3.6 Earned Business Value

In traditional projects, project managers often use EVA (Earn Value Analysis) metric to receive a clearer picture of a project's progress. By using EVA, they really want to know how much value the product is currently providing or what percentage of the product is "done." EVA in standard definition relies on budgeted costs [35] which are unavailable in agile, because of lack of up-front high weight planning. Instead of EVA, EBV (Earn business Value) is used in agile and mainly used for tracking the "valuable" part of "valuable software." This metric can be measured in terms of financial value that is based on the estimated ROI prorated to features of each User Story. Also, EBV measures the extent to which a product is complete, from a business perspective [36].

EBV in agile focuses on business value and hence, no big up-front design is needed. For calculating EBV in a project, manager breaks down the project based on the extracted and defined features and user stories. Each story has its own relative weight assigned by product owner. Rawsthorne defined a formula for calculating EBV of each story[36] as the sum of weight for stories done. He also discussed on this metric in Scrum in another article [37].

It can be said, this metric is useful only when scope of the work is clear up-front. But if level of uncertainty related to the scope is high, it is completely useless [38-40].

3.7 Total Effort Estimation

In formal and ideal view, agile teams mostly prefer to predict and undertake in doing specific USs only for the next iteration [17]. But in the real world, managers require an up-front prediction for all of the work in order to dedicate the necessary budget. An agile team, based on its velocity and the predicted USPs for all of the backlog items is able to guess which USs can be provided in the future iteration and also how many iterations need for completing the project [2]. An initial prediction assumes that large number of items will not change within the project such as velocity or factors impressing the velocity which have been specified from previous projects and current amount of the product backlog. These factors are team combination, selected technology and framework, development process and non functional requirements.

Usually, estimating of project duration is based on the number of iterations of a constant and fixed time [2]. Total Effort estimation is based on the number of dedicated and full time team members on the project [14, 15]. Of course, this approach provides many weaknesses, for example when some individuals or skills are required part-time and also when many low or average priority items are considered at wrong level so estimation becomes hard in USPs [3].

3.8 Compare and Summarize

As explained, there are different approaches and practices in measuring values in ASD methods. Each of them is used for specific aim and in specific time. Also, these practices have different basis and sometimes have different targets and goals. In table 1 we present main aim and basis of each practice. It can be said that each of the practice focuses on one the agile value and so, in prospective of that value, respected practice is useful and play a significant role.

In addition, each practice is helpful for specific team member(s) and also can be used or done in particular time. Note that some of the practices would be used many times within the project. Table 2 illustrates these related team member(s) and frequency of each practice.

Table 1: Main Aim and basis of practices

| Practice | Main aim | Basis of practice |
|---|---|---|
| Software size | Estimation of software size/effort | User Stories |
| Velocity | Overall productivity of team | User Story points |
| Burndown chart | Progress monitoring | User Stories |
| Cumulative flow | observation of lead time and WIP queue depth | Work in Process/Progress |
| Responding to change | indicator of ability of team to hand over product quality | Defects fixing cost |
| Earned Business Value | Monitoring business value delivered to customer | Business value |
| Total Estimation Effort | Planning and budgeting | User Stories and Re-works |

Table 2: Main Aim and basis of practices

| Practice | Beneficiary | Frequency |
|---|---|---|
| Software size | Project manager | At start of each iteration |
| Velocity | Project manager | At end of each iteration |
| Burndown chart | All team members | At end of each iteration |
| Cumulative flow | Top managers /customers | At end of each iteration |
| Responding to change | project manager | At end of each iteration/ At end of project |
| Earned Business Value | Top managers /customers | As each feature is delivered |
| Total Estimation Effort | Top managers /customers | At beginning of the project |

All of the above practices are used in ASD methods and almost none of the traditional measurements are usable directly in these methods. Only COSMIC published an adopted version of COSMIC function point [47] which in the next section we have focused on it.

## 4. COSMIC for Agile

Always software size estimation has been attractive for computer scientist. Albrecht's work [41] on measuring software size was the first acceptable practice for this matter. After that, many methods and techniques were published based on software functionality. Software function measuring (FSM) is now widely used in software industry for many reasons, but the most important one is as input for effort and cost estimation models. The most popular FSM methods are FPA [41], De Marco's bang model [42], Mark II [43], Boeing 3D function point [44], Full Function Point [45], COSMIC FFP [46]. Some of these methods are adopted by ISO/IEC 14143 and now are accepted as international standard for FSM. Unfortunately, most of them are not usable in ASD methods since they need well defined requirements.

Among all different afore mentioned methods, COSMIC FFP or CFP seems more applicable in agile projects, mainly because CFP does not require detailed specifications and also, mapping between CFP and USP is easy and understandable. CFP method measure functional user requirement and estimation is derived in terms understood by users of the software. In CFP four base functional components types (read, write, entry and exit) are extracted and estimation relies on these items [47].

First attempt to use COSMIC in agile project was in 2011, by Desharnais and et al. [48] They proposed a procedure for using CFP in agile methods and assessed it in a real project. However, there is still a little amount of guess estimation on some of the USs, but by eliciting requirements from USs and focusing on high quality of documentation of USs, this methods is helpful.

In late 2011 based on the Desharnais' work, COSMIC published officially an agile version of COSMIC FFP for using in agile software development [3]. In agile COSMIC, according to the agile features, some modifications were made. Each User Story is defined as a single FP. To satisfy requirement changes requested by customer, functional size of each User Story could include changes to a previously released User Story.[3] By this trick, any change requirement among the next iteration could be calculated and hence, based on adding, changing or cancellation of any data movement, size of software could be changed. Actually, estimation is done in start of iterations and updates previous predictions. It is clear that COSMIC method is used for size measurement only and not for effort estimation directly.

It seems that these minor modifications on standard COSMIC for adapting to agile projects cause this method being useful for managers and stakeholders in agile environment.

Using CFP in agile methods leads to achieving objective and repeatable measurement. As mentioned in previous section, poker planning method is completely subjective and based on average USPs. In addition, using CFP software size indicator "should take no more effort than in units of USP" [3].

When clients need to use a standard method for estimation in agile project using CFP as a standard will be helpful.

## 5. Summary and future work

Measurement in agile software development methods is different from traditional methods. This diversity is mostly because of different process in these methods. While one emphasizes on documentation, another focuses on lightweight documents. A few metrics and measurements are defined especially for agile methods, which some of them are easy to use and useful. Using velocity measurement in stable teams and in long periods of time is a helpful metric for overall team productivity. Software size estimation in agile methods is team-driven measurement and despite the traditional methods, size estimation process is done for any iteration separately. Burndown and burn up charts also are usable for monitoring and controlling project progress and are popular in all agile teams. EBV is another significant measurement in agile teams. However, this metric is usable when scope of project is well-defined in advance; it seems that it is more helpful for small and medium agile projects.

Although, there are few efforts for agile measurement, because of permissibility of changing requirement, achieving a comprehensive measurement method in all activities and process is so difficult. COSMIC in its first effort, published a guideline for estimation of software size in agile projects. Using a standard method for size estimation in agile development could be useful. Agile COSMIC is based on USs and simply, size estimation in first of iteration is based on user stories. This estimation in next iteration will be updated in case of adding, changing or cancelling any function. This is a good practice for size estimation in agile, because this method is consistent with agile principles and does not need to detail specifications, also, because of introducing an objective measurement instead of popular subjective practices.

As we explained, only COSMIC published a measurement guideline for ASD methods and in this guideline only software size will be estimated. Yet there is no well-defined and standard other measurement practices in ASD methods. For the future work, there are some hot topics in this area such as measurement of productivity and not velocity. Also change requirement measurement is another issue that can be studied later.

## Acknowledgments

We would like to thank to Mr. Reza Meimandi Parizi for reviewing this work and valuable guidance and advice.

**Taghi Javdani Gandomani** has 10 years experience in both industry and academic research in software methodologies and project management. He obtained Ms. of Software Engineering from Isfahan University in Iran in 2001. Currently, he is now PhD student in University Putra Malaysia and is working on agile methodologies.

**Hazura Zulzalil** holds a Ph.D. from University Putra Malaysia. Currently, she is a senior lecturer at the Faculty of Computer Science and Information Technology, University Putra Malaysia. Her research interests are software metrics, software quality and software product evaluation.

**Abdul Azim Abd Ghani** obtained his Ph.D. from University of Strathclyde. Currently, he is a Professor at the Faculty of Computer Science and Information Technology, University Putra Malaysia. His research interests are software engineering, software measurement, software quality, and security in computing.

**Abu Bakar Md. Sultan** holds a Ph.D. from University Putra Malaysia. Currently, he is an Associate Professor and the Head of Information Systems Dept., Faculty of Computer Science and Information Technology, University Putra Malaysia. His fields of expertise are Metaheuristic and Evolutionary Computing.